\documentclass[twocolumn]{aastex631}

\usepackage{amsmath}

\begin{document}

\title{Collisionless tearing instability in relativistic non-thermal pair plasma and its application to MHD turbulence}

\author[0000-0002-9217-6964]{Ivan Demidov}
\affiliation{Physics Department, Ben-Gurion University, PO Box 653, Beer-Sheva 84105, Israel}

\author[0000-0002-2316-8899]{Yuri Lyubarsky}
\affiliation{Physics Department, Ben-Gurion University, PO Box 653, Beer-Sheva 84105, Israel}



\begin{abstract}

Collisionless tearing instability with a power-law distribution function in a relativistic pair plasma with a guide field is studied. When the current sheet is supported by plasma pressure, the tearing mode is suppressed as the particle spectrum hardens. In the force-free limit, the instability growth rate becomes independent of the particle spectrum. We apply these results to relativistic MHD turbulence, where magnetic energy greatly exceeds plasma rest energy, and derive an expression for the transverse size of turbulent eddies unstable to tearing mode. We also establish the critical plasma magnetization parameter above which charge starvation prevents the tearing instability. These results might be useful for developing more accurate models of particle acceleration in relativistic astrophysical sources.

\end{abstract}



\section{Introduction} \label{sec:intro}

Magnetic reconnection is a fundamental process in plasma physics, where magnetic field lines in a highly conductive plasma converge, break, and reconnect changing their topology. This process leads to a rapid conversion of magnetic energy into bulk kinetic energy, thermal energy, and particle acceleration. It plays a crucial role in various astrophysical plasmas, such as solar flares \citep{Innes2015}, pulsar winds and nebulae \citep{Coronity1990,Lyubarsky2001},  jets from active galactic nuclei \citep{Romanova1992,Christie2019}, gamma-ray bursts \citep{McKinney2012,Lazarian2019} and black hole magnetospheres \citep{Bransgrove2021,Ripperda2022}.

Reconnection is important in the context of the MHD turbulence which is a ubiquitous feature of astrophysical plasmas.  
Several investigations have proposed that turbulent fluctuations in non-relativistic magnetized plasmas can lead to the formation of elongated small-scale current sheets (\citealt{Boldyrev2005}) that can be disrupted via magnetic reconnection, providing an additional mechanism for energy dissipation and particle acceleration (\citealt{Uzdensky2016,Mallet2017,Loureiro2017}). 
Additionally, the intermittency inherent in MHD turbulence leads to the formation of large-scale sheet-like structures, which can also be unstable with respect to plasmoid formation (\citealt{Mallet2017}, \citealt{Dong2018}, \citealt{Chernoglazov2021}).  In turbulent magnetized plasmas, reconnection initiates the "injection" phase of particle acceleration, where particles gain energy from non-ideal electric fields within current sheets. This is followed by a "stochastic" acceleration stage, mediated by scattering off turbulent plasma fluctuations \citep{Comisso2019}. According to particle-in-cell (PIC) simulations, stochastic acceleration allows particles to reach Lorentz factors where their Larmor radius is of the order of the turbulence driving scale. However, in realistic astrophysical contexts, this scale is so huge that achieving such high Lorentz factors is impractical due to significant radiative cooling effects. As demonstrated by \cite{Nattila2021}, radiative cooling suppresses stochastic acceleration and strongly affects the maximum particle energy. Understanding the connection between MHD turbulence and particle acceleration is essential for explaining the high-energy radiation observed in relativistic astrophysical sources; for this reason, so much research is devoted to this topic (e.g., see some recent works \citealt{Vega2024b}, \citealt{Lemoine2024}, \citealt{XuLazarian2023}, \citealt{Vega2022a}).

In the above studies of MHD turbulence, the primary mechanism leading to magnetic reconnection is the tearing instability. This process occurs when a plasma current sheet becomes unstable, resulting in the formation of magnetic islands and reconnection of magnetic field lines (\citealt{Furth1963}, \citealt{Laval1966}, \citealt{Coppi1966}).
Even though non-relativistic collisionless tearing instability has been extensively investigated, its relativistic counterpart remains less explored. \cite{Zelenyi1979} considered the collisionless tearing mode with the guide field in the relativistic pair plasma analytically by using the kinetic approach. This approach was also used in \citealt{Holdren1970}, \citealt{Petri2007}, \citealt{Zenitani2007}, \citealt{Hoshino2020}, where analytical estimates were also confirmed by numerical calculations. Maxwellian plasma was assumed in all of the aforementioned studies. Collisionless tearing instability in the relativistic pair plasma with the non-Maxwellian distribution function of particles was recently studied by \cite{Thompson2022}. He considered specific conditions in the pulsar emission zone, namely, weakly sheared quantizing magnetic field and the narrow top hat distribution function centered at characteristic momentum $p_0$ both for electrons and positrons. 
Relativistic tearing instability with the guide field was studied in the fluid approach by \citet{Lyutikov2003, Komissarov2007, Yang2019}. Despite the simplicity of the fluid equations, there is the closure problem for the plasma pressure tensor in collisionless plasmas and kinetic theory is needed to determine effective resistivity, which is often considered as a free parameter.


In relativistic astrophysical sources, plasma could hardly reach the Maxwellian distribution since binary particle collisions are typically too rare due to low densities and high temperatures. 
Power-law particle spectrum is a common feature observed in such high-energy
environments. It is characterized by a functional form $\mathcal{N}(\mathcal{E})\sim \mathcal{E}^{-\alpha}$, where 
$\mathcal{N}(\mathcal{E})$ is the number of particles with energy $\mathcal{E}$ and $\alpha$ is the particle spectral index. 
In our previous paper \cite{DemidovLyubarsky2024}, we examined the tearing instability in ultrarelativistic pair plasma with a power-law distribution function in the absence of the guide field. We analytically estimated the growth rate and validated these results by numerically solving the dispersion equation, accounting for complex particle trajectories within the reconnecting layer. We found that the instability is suppressed when the particle spectrum becomes harder. Indeed, the growth rate is determined by the ratio of the thickness $l\sim \sqrt{2r_g L}$ of the resonant layer, where the particles gain energy from the induced electric field to the thickness $L$ of the current sheet itself. Since in the relativistic case, the effective mass of particles depends on their energy, the resonance layer thickness $l$ is always determined by the Larmor radius of low-energy particles.
In turn, the minimal thickness of the current sheet $L$ depends on the slope of the spectrum. At $\alpha \geq 2$, thermal pressure is dominated by low-energy particles that make possible $L_{\text{min}}\sim l$. At $\alpha < 2$, thermal pressure and current sheet thickness $L$ are determined by a small fraction of high-energy particles such that $L$ cannot be smaller than the Larmor radius of high-energy particles, which leads to $L_{\text{min}} \gg l$ and significantly suppresses the tearing mode.

In this paper, we investigate how a non-thermal particle spectrum with wide energy spread affects the collisionless tearing mode in the ultrarelativistic pair plasma in the presence of the guide field. We consider two qualitatively different situations: 1) the uniform guide field and  2) the force-free limit, where plasma pressure is negligible, such that it is necessary to assume that the guide field depends on coordinates to maintain the equilibrium of the current sheet. We obtain the estimation of the tearing instability growth rate in the linear regime and find how it depends on the power-law spectral index, $\alpha$. Then, we use the resulting expression for the growth rate in the force-free limit to understand whether tearing instability can affect the spectrum of relativistic MHD turbulence when magnetic energy density greatly exceeds the plasma rest energy.

The paper is organized as follows: In Sections 2 and 3 we obtain analytical estimates for the growth rate of tearing instability both for the uniform guide field and the force-free limit, respectively. In Section 4, we determine the conditions under which tearing-mediated MHD turbulence is possible. In Section 5 we explore various astrophysical applications.
Finally, in Section 6, we provide the discussion and summary of our results.

\begin{figure}
  \centering
  \includegraphics[width=\columnwidth]{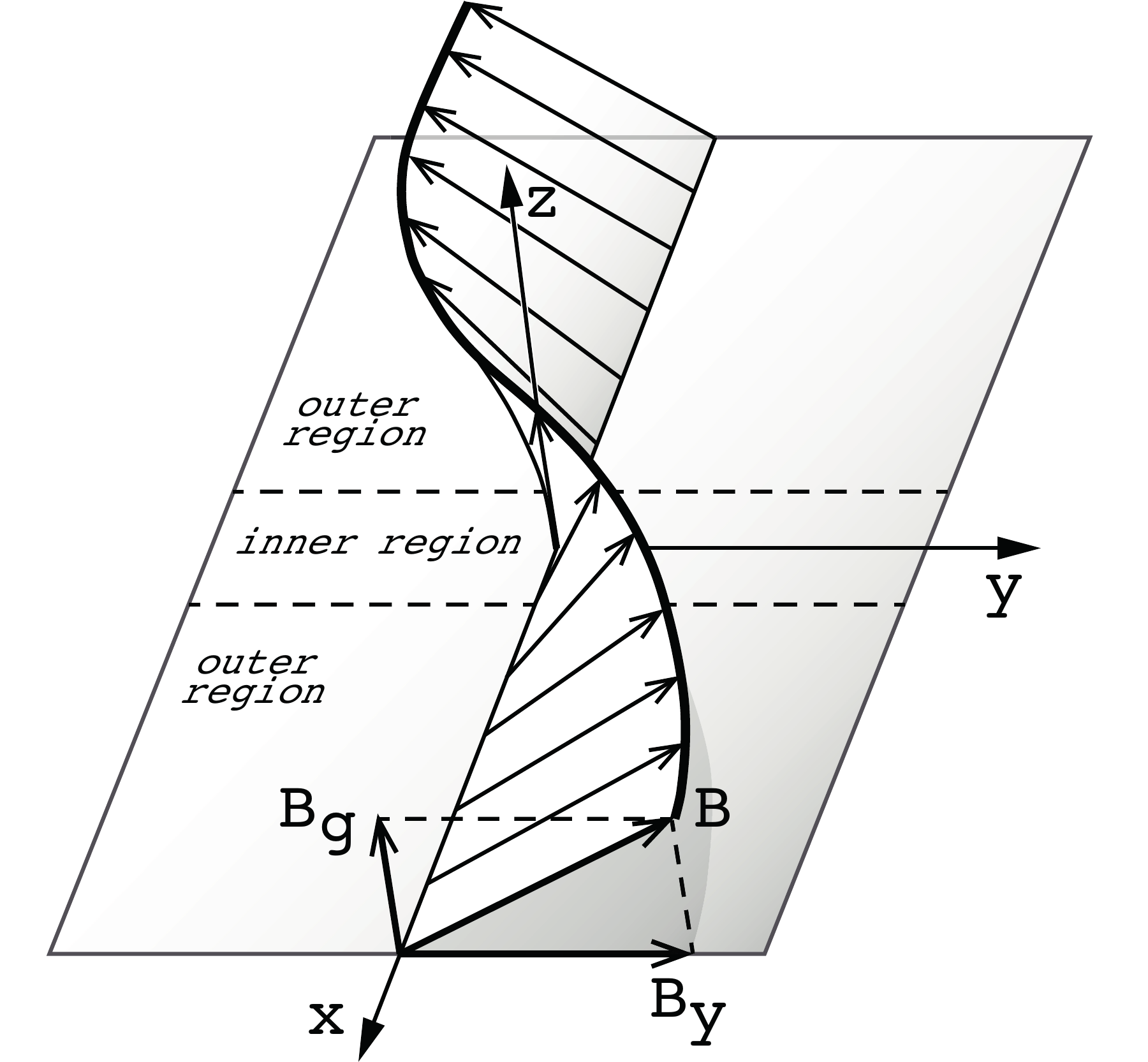}
  \caption{Unperturbed magnetic field configuration within the current sheet with a guide field}
\label{f1}
\end{figure}

\section{Tearing instability in the case of the uniform guide field}\label{sec:uniform}

\subsection{The current sheet equilibrium}

Let the unperturbed magnetic field be $\mathbf{B}=B_{0y}f(x)\hat{\mathbf{y}}+B_g\hat{\mathbf{z}}$, such that the reconnecting magnetic field with a profile $f(x)$ is directed along $y$-axis and the uniform guide field $B_g$ is directed along $z$-axis. 
Thus, across the current sheet, the total magnetic field $\mathbf{B}$ rotates through a finite angle less than $180^\circ$ (see Fig.~\ref{f1}). It is assumed that 
\begin{equation}\label{str}
B_g> B_{0y}\sqrt{r_{0g}/L},
\end{equation}
where $r_{0g}=\mathcal{E}/eB_{0y}$ and $L$ is the current sheet thickness, i.e. plasma particles are fully magnetized. This inequality shows that the Larmor radius $r_g=\mathcal{E}/e B_g$ in the guide field is less than the thickness of the layer $ l\sim\sqrt{2r_{0g}L}$ near the plane $x=0$, where particle would be unmagnetized and perform complicated meandering motion in the absence of the guide field \citep{Parker1957}. If the spectrum of particles satisfies the inequality $\alpha<2$, such that most of the plasma energy is contained in high-energy particles, there are two energy scales $\langle\mathcal{E}\rangle\sim \mathcal{E}_{\text{max}}$ and $\langle\mathcal{E}^{-1}\rangle^{-1}\sim \mathcal{E}_{\text{min}}$. Thus, it leads to three possible cases: 1) particles of all energies are not magnetized by the guide field; 2) only particles of low energy $\sim \mathcal{E}_{\text{min}}$ are magnetized; and 3) particles of all energies are magnetized. The first case reduces to the problem of tearing instability in the case of the usual Harris equilibrium. The second case does not differ quantitatively from the third one because unmagnetized high-energy particles make a small contribution to the perturbation of the total current density (see Appendix A). Therefore, we consider the third case only, i.e. when $B_g>B_{0y}\sqrt{\mathcal{E}_{\text{max}}/eB_{0y}L}$. 

It is assumed that electrons and positrons perform the macroscopic drift in the $z$-direction at a velocity $\pm U$ ("$+$" for positrons and "$-$" for electrons, $U>0$), such that they create 
the current density that ensures the variation of the reconnection magnetic field $B_y$ with $x$ (see Appendix B for details). 

Since $B_g=\text{const}$, the equations describing the current sheet equilibrium are the same as in the case without the guide field: 
\begin{equation}\label{two}
    \frac{B_{0y}^2}{8\pi}=\frac{2}{3} n_0\langle\mathcal{E}\rangle, \quad \frac{U}{c}\sim \frac{\langle\mathcal{E}\rangle}{eB_{0y} L},
\end{equation}
where $\langle\mathcal{E}\rangle$ is the average energy of particles, $n_0$ is the equilibrium number density at $x=0$ of one kind of particles. The first equation is the pressure balance and the second one is just the definition of the characteristic diamagnetic drift velocity of the plasma. 

\begin{figure}
  \centering
  \includegraphics[width=\columnwidth]{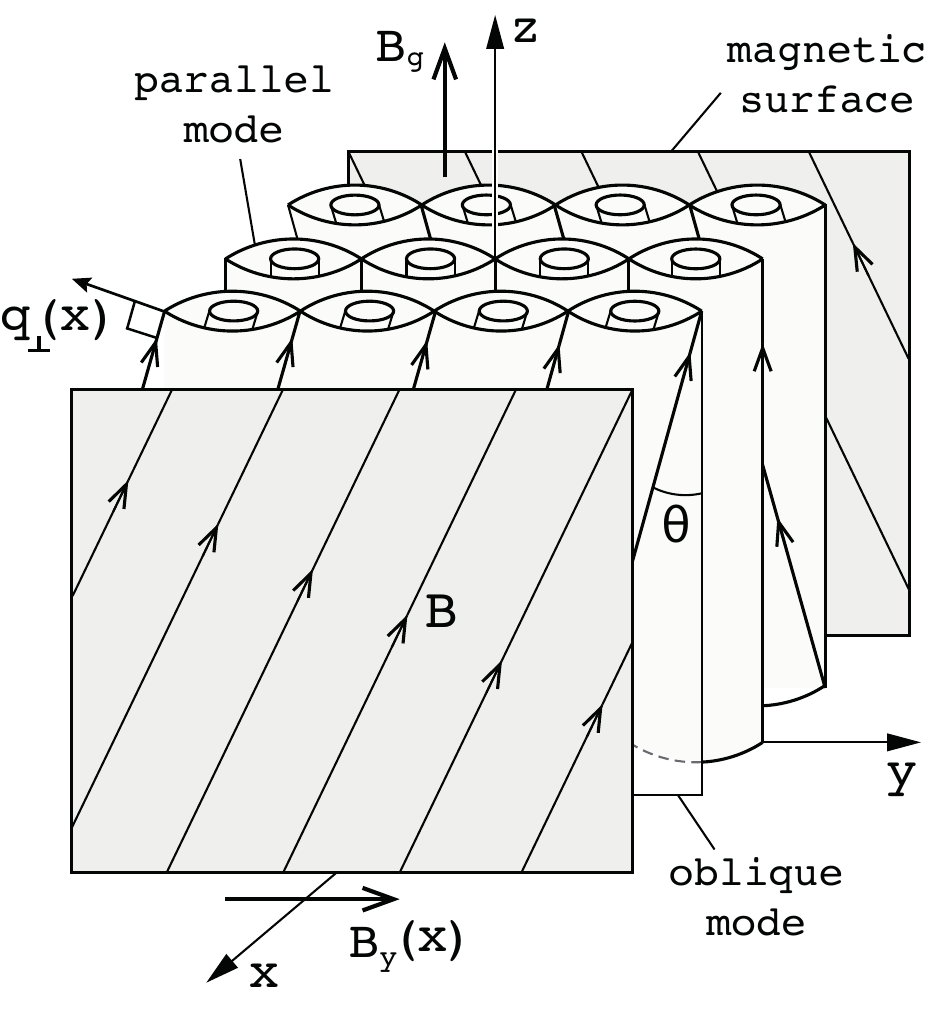}
  \caption{Perturbed magnetic field configuration within the current sheet}
\label{f2}
\end{figure}

The magnetic field configuration shown in Fig.~\ref{f1} can be considered as a set of magnetic surfaces parallel to the $x=0$ plane. Within each such surface the total magnetic field $\mathbf{B}$ is rotated at a certain angle $\theta(x)$ relative to the $z$-axis. This system with a current sheet is unstable to pinching; it leads to reconnection of the magnetic field and formation of magnetic islands (see Fig.~\ref{f2}). In fact, this is possible only if the resistivity of plasma is not zero. In the considered case of collisionless plasma with the guide field, the resistivity is provided by the inertia of charge carriers. 

It is assumed that the time and coordinate dependence of the perturbed magnetic field has the following form
\begin{equation}\label{6-03}
\delta\mathbf{B}=\delta\mathbf{B}_1(x)\exp\left(-\text{i}\omega t+\text{i}\boldsymbol{q}\cdot\boldsymbol{r}\right).
\end{equation}
The pinching of the current is the most effective in a direction perpendicular to the magnetic field within the same magnetic surface (e.g., \citealt{Galeev1984}). Thus, perturbations of the tearing mode satisfy the following condition (see also Section 2.3)
\begin{equation}
q_\parallel(x)=\frac{\mathbf{q}\cdot\mathbf{B}(x)}{B(x)}\approx 0
\end{equation}
and can be described by the transverse component of the wave vector $\mathbf{q}_\perp=q_y\hat{\mathbf{y}}+q_z\hat{\mathbf{z}}$. This means that we have multiple modes with different $\mathbf{q}_\perp$, and 
different orientations of $\mathbf{q}_\perp$ corresponds to different magnetic surfaces with coordinates $x_s=-L\arctan[q_z B_{z}/(q_y B_{0y})]$ where the tearing mode is occur (see Fig.~\ref{f2}, where three surfaces with tearing mode are shown). 

In the strong guide field regime $B_g>B_{0y}$, the fastest-growing instability is the oblique tearing mode at $\theta\sim\theta_{\text{max}}=\arctan(B_{0y}/B_g)$. However, the deviation from the growth rate at $\theta = 0^\circ$ does not exceed a factor of two (\citealt{Liu2013}, \citealt{Akcay2016}). 
Since we are limited to order-of-magnitude estimations and investigate the effect of the particle spectrum on the linear stage of tearing instability, we consider only parallel perturbation with the wave vector $\mathbf{q}=q_y\hat{\mathbf{y}}$ that corresponds to the magnetic surface with the location $x_s=0$. Consideration of other modes in pair plasma might be important at the nonlinear stage and relatively weak guide fields when magnetic islands on different magnetic surfaces overlap and lead to stochastic field lines and a turbulent plasma evolution (\citealt{Galeev1984}, \citealt{Daughton2011}).



Slow perturbations far from the plane $x=0$, where the electric field is absent, could be described in the ideal MHD approximation. 
For simplicity, we consider the current sheet with the magnetic field profile $f(x)=\tanh(x/L)$ \citep{Harris1962}. Near the plane $x=0$, the MHD approximation is violated, and kinetic effects are important. Thus, one can find a solution in the inner and outer regions and then match them together, which ultimately gives an expression for the growth rate of tearing instability.

\subsection{The outer region}

Slow plasma motions due to the tearing instability create the current density perturbation, which, in turn, creates a disturbance of the magnetic field $\delta\mathbf{B}(x,y)$ that can be expressed via the vector potential perturbation according to $\delta\mathbf{B}=\text{rot}\,\delta \mathbf{A}$.
Assuming the reconnecting magnetic field profile $f(x)=\tanh(x/L)$, the vector potential perturbation in the outer layer is well-known (e.g. \citealt{Sturrock1994}) and can be written as
\begin{equation}\label{outer}
\delta A_z(x)=\delta A_z(0)\left(1+\frac{1}{q L}\tanh\frac{|x|}{L}\right)\exp\left(-q|x|\right),
\end{equation}
where $q=|q_y|$ and the even-parity solution of Ampère's equation is chosen. The odd-parity solution $\delta A_z(-x)=-\delta A_z(x)$ corresponds to the kink mode, which is stabilized by the guide field and is not interesting to us \citep{Zenitani2008}.

It is clearly seen from the solution~(\ref{outer}) that the derivative $\delta A'_z(x)$ is discontinuous in the general case. Let us introduce the notation
\begin{equation}\label{delta}
    \Delta'(0)=\frac{\delta A'_z(0+)-\delta A'_z(0-)}{\delta A_z(0)}.
\end{equation}
According to~(\ref{outer}), we obtain
\begin{equation}\label{Hdelta}
    \Delta'(0)=\frac{2(1-q^2L^2)}{q L^2}.
\end{equation}
Therefore, $\Delta'(0)=0$ and there is no discontinuity $[\![\delta B_y]\!]=0$ if and only if $q L= 1$.
At $q L\neq 1$ the magnetic field perturbation jump occurs. To remove the discontinuity of the magnetic field, we have to take into account the current of particles within the inner region $|x|<l$.

\subsection{The inner region}

For the inner region, we can use the same line of reasoning that was used by \citealt{DrakeLee1977}, with the only difference that we are considering relativistic particles. It was shown by \cite{Baalrud2018} that the analytical solution of Drake and Lee theory accurately predicts the growth rate of parallel ($\theta=0^\circ$) modes in non-relativistic pair plasma if the current sheet is thick $L\gg r_g$. 
When $L\sim r_g$ the theory gives correct results by order of magnitude. 
Thus, the chosen Drake and Lee approach is quite reliable.
Our goal is to calculate the current density perturbation $\delta j_z$ in the inner region.

Since the total vector potential has only $z$-component as in the case without the guide field, reconnection induces an electric field $\delta E_z=\text{i}(\omega/c)\delta A_z$ directed along the current sheet. This electric field is periodic along the $y$-axis; it is directed in negative $z$-direction in the vicinity of X-points and in positive $z$-direction between the  X-points. However, magnetized particles cannot be accelerated strictly along the electric field since they are allowed to move 
mostly along the total magnetic field $\mathbf{B}$ directed at an angle $\theta(x)$ to the $z$-axis. 
\begin{figure}
  \centering
  \includegraphics[width=\columnwidth]{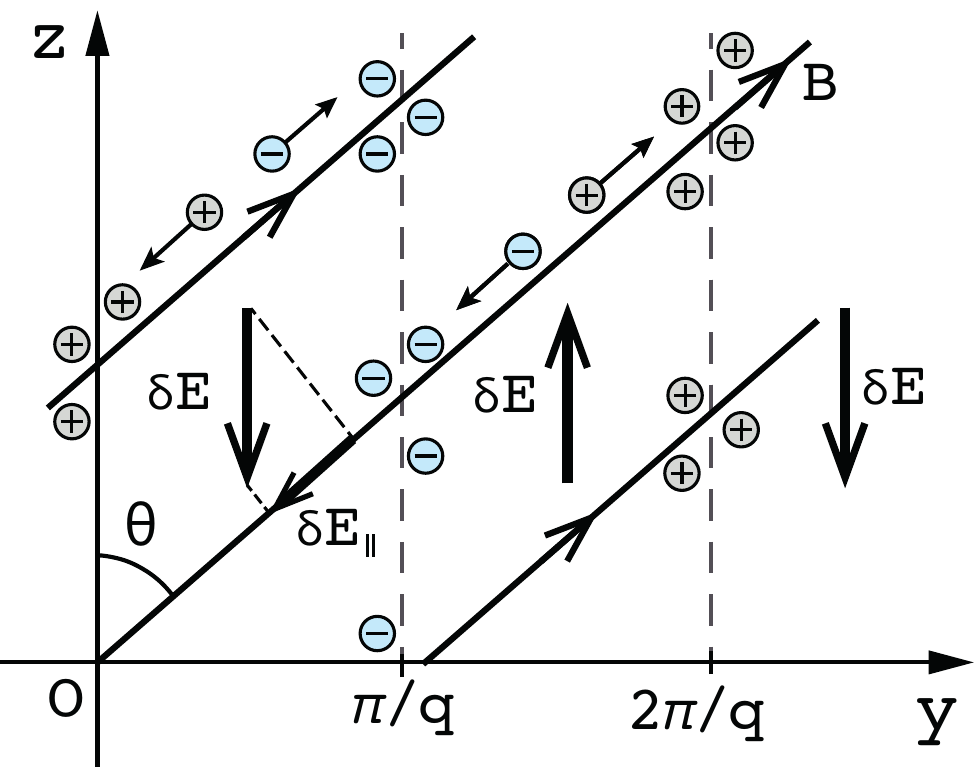}
  \caption{Electric field stripes and charge separation of low-energy particles. This separation neutralizes the electric field $\delta E_\parallel$ and suppresses the tearing mode.}
\label{f5}
\end{figure}
While the particles move along the total magnetic field, they "see" a rapidly varying electric field $\delta E_\parallel=\delta E_z\cos\theta$, since motion along the inclined magnetic field line also assumes motion along $y$-axis, where the periodicity of $\delta E_z$ plays a role. Therefore, the average acceleration of the particles is small when the magnetic field line crosses many periods of the electric field, i.e. when $q_\parallel =q_y\sin\theta$ is not small. Moreover, the motion of low-energy particles along $\mathbf{B}$ leads to charge separation. Indeed, within the “stripe” of the electric field of the same polarity, electrons and positrons move along $\mathbf{B}$ in different directions, thereby accumulating at the boundaries between the adjacent "stripes" (see Fig.~\ref{f5}).
It leads to an electrostatic field $-\text{i}q_\parallel\delta\phi$ generation, which tends to compensate the induced parallel electric field $\delta E_\parallel$. Unless $q_\parallel$ is small enough, the total electric field along $\mathbf{B}$ is practically zero, and no current perturbation $\delta\mathbf{j}$ arises. Thus, the only possibility for the development of the instability is $q_\parallel(x)\approx 0$ which is possible only in the narrow layer $|x|<l$ when the total magnetic field is directed almost along $z$-axis. Roughly, it suggests that the magnetic field line along which the particles move is largely confined within a region of uniform electric field polarity.

Let us estimate the thickness $l$ of the resonance layer. This value is regulated by the angle of rotation of the magnetic field line $\theta(x)$. Indeed, if the particle acceleration time is of the order of $1/\gamma$ (where $\gamma$ is the tearing instability growth rate), and the thickness of the electric field "stripe" is of the order of $1/q_y$, then the angle of rotation of the magnetic field $\theta$ should not exceed $\gamma/q_yc$. Only in this case, particles move within only one "stripe" of the electric field during the instability. On the other hand, the angle between the total magnetic field and the $z$-axis is $\theta\sim B_{0y}(l/L)/B_g$. Therefore, the thickness $l$ can be estimated as
\begin{equation}\label{region}
    l\sim \frac{\gamma}{q_y c}\frac{B_g}{B_{0y}}L.
\end{equation}
We assume that particles of all energies are magnetized and their motion can be approximated as one-dimensional, like beads on a wire.
This is possible only
when the Larmor radius $r_g$ in the guide field is much smaller than the thickness of the inner region $l$. However, as was shown by \cite{Baalrud2018}, the small gyroradius expansion is still valid even then $r_g>l$. In fact, the resonance layer is smeared over a few Larmor radii since the magnetic field cannot change over a distance smaller than the Larmor radius. Given that the magnetic field jump, 
$[\![\delta B_y]\!]$, is determined by the outer region, the total current must remain the same.  Thus, increasing the resonance layer thickness leads to a decrease in the current density and both these factors cancel each other.  
Therefore, one can use the one-dimensional approximation; then the equation of particle motion is
\begin{equation}\label{eqnmotion}
    \frac{\text{d}}{\text{d} t}(\Gamma m \delta v_z)=Q_s \delta E_z,
\end{equation}
where $\Gamma$ is the Lorentz factor of the particle and $Q_s$ is the particle charge.

Since the electric field can be written in the form $\delta E_z=-(1/c)\text{d}(\delta A_z)/\text{d}t$, we can integrate the equation of motion over $t$, which gives $\delta v_z\approx -(Q_s/\Gamma m c)\delta A_z$. The corresponding current density is
\begin{equation}\label{curr}
    \delta j_z=\sum_s Q_s n_s \delta v_z\approx \frac{2n_0 e^2}{\Gamma m \gamma}\delta E_z.
\end{equation}
The coefficient in front of the electric field resembles the Drude formula for the electrical conductivity coefficient, where the characteristic time of "collisions" is equal to the inverse growth rate of the instability, i.e. $\tau\sim \gamma^{-1}$. Another way to look at this relationship is to note that $\gamma\delta j_z\sim \text{d}(\delta j_z)/\text{d}t$. This term enters the generalized Ohm law in the fluid description of tearing instability (e.g. \citealt{Vasyliunas1975}) and characterizes the violation of frozen-in condition due to the inertia of charge carriers.

Since we consider an ensemble of particles, we should write the average inverse Lorentz factor $\langle mc^2/\mathcal{E}\rangle$ instead of $1/\Gamma$. Therefore,
\begin{equation}\label{current}
    \delta j_z \approx \frac{2n_0 e^2}{m\gamma}\Big\langle\frac{m c^2}{\mathcal{E}}\Big\rangle \delta E_z.
\end{equation}
It can be seen that particles with the lowest energy mostly contribute to the conductivity.


\subsection{The growth rate}\label{subsec:rate}

The current perturbation $\delta j_z$ 
removes the discontinuity of the magnetic field perturbation. According to~(\ref{current}), the Ampère's law,
\begin{equation}\label{amp}
     [\![\delta B_y]\!]\approx \frac{4\pi}{c}\delta j_z l,
\end{equation}
can be rewritten as
\begin{equation}\label{guid}
    \Delta'\approx 8\pi n_0 e^2\langle\mathcal{E}^{-1}\rangle l,
\end{equation}
where $\Delta'=\Delta'(0)$ is determined by~(\ref{delta}). Combining~(\ref{region}) and~(\ref{guid}) with the expression $\Delta'=2(1-q^2L^2)/q L^2$ for the case of the Harris sheet, we obtain
\begin{equation}\label{inc}
    \gamma(q)\sim \frac{c}{L^3}(1-q^2L^2)\frac{B_{0y}}{B_g}\frac{\langle\mathcal{E}^{-1}\rangle^{-1}}{4\pi n_0 e^2}.
\end{equation}
To eliminate the particle density one can use equations for the Harris equilibrium~(\ref{two}).
It gives the following estimation for the growth rate
\begin{equation}\label{tearing_incr}
    \gamma(q)\sim \frac{c}{L}(1-q^2L^2)\frac{B_{0y}}{B_g}\left(\frac{U}{c}\right)^2\left[\langle\mathcal{E}\rangle\langle\mathcal{E}^{-1}\rangle\right]^{-1}.
\end{equation}
In general case, characteristic energies $\langle \mathcal{E}\rangle$ and $1/\langle \mathcal{E}^{-1}\rangle$ may not coincide. At large $\alpha\geq 2$, we expect $\langle \mathcal{E}\rangle\sim\langle\mathcal{E}^{-1}\rangle^{-1}\sim\mathcal{E}_{\text{min}}$ and the growth rates are determined by the same formula as for the relativistic Maxwellian plasma with $\mathcal{E}_\text{min}\sim k_B T$ \citep{Zelenyi1979}. In the case of the power-law distribution with $\alpha<2$, one can expect that $\langle \mathcal{E}\rangle\sim \mathcal{E}_{\text{max}}$ and $\langle \mathcal{E}^{-1}\rangle^{-1}\sim \mathcal{E}_{\text{min}}$, i.e., there are two characteristic energy scales. Then, the growth rate contains a small factor $\epsilon$, where
\begin{equation}
\epsilon=\frac{\mathcal{E}_{\text {min}}}{\mathcal{E}_{\text{max}}}\ll1.
\end{equation}
The obtained expression for the growth rate is valid in the limit $1-qL\ll 1$, when the constant-$\delta A$ approximation is valid, i.e. when the electric field is not changed strongly across the inner region. 
In general, the obtained growth rate differs from the non-relativistic analog only by the particle velocity ($v\approx c$), as well as by additional factors that take into account relativistic effects and the non-Maxwellian spectrum. These factors do not affect the dependence on the wave number $q$. Therefore, as in the non-relativistic case, the maximum growth rate is achieved at $q\sim 1/2L$ (e.g., \citealt{Daughton2005}, \citealt{Baalrud2018}), and
\begin{equation}
    \gamma_{\text{max}}\sim \frac{c}{L}\frac{B_{0y}}{B_g}\left(\frac{U}{c}\right)^2\left[\langle\mathcal{E}\rangle\langle\mathcal{E}^{-1}\rangle\right]^{-1}.
\end{equation}



\subsection{Dependence on the power-law index}
Let us find how the growth rate depends on the power law index of the particle spectrum, $\alpha$. 
It is worth comparing only the maximum growth rates when $U\sim c$ and the current density is maximum, $j_{\text{max}}=2en_0c$. Formally, this condition also means that the average Larmor radius in the magnetic field $B_{0y}$ is comparable with the current sheet thickness $L$ (see equation~(\ref{two})). Even though our calculations assumed $U\ll c$, the obtained results could be used as an order of magnitude estimation even at $U\sim c$. 

Since for $\alpha\geq 2$, there is only one energy scale $\mathcal{E}\sim\mathcal{E}_{\text{min}}$ as in the Maxwellian case, the growth rate $\gamma(\alpha\geq2)$ is of the order of that for the Maxwellian distribution function with the effective temperature $k_B T\sim\mathcal{E}_{\text{min}}$. In this case, we should use $L\sim \mathcal{E}_{\text{min}}/eB_{0y}$. At $\alpha<2$, the minimal width of the current sheet is of the order of the Larmor radius of the energetic particles, $L\sim \mathcal{E}_{\text{max}}/eB_{0y}$. Assuming that the magnetic field $B_{0y}$ is the fixed external parameter, we obtain
\begin{equation}\label{ratio}
    \frac{\gamma(\alpha<2)}{\gamma(\alpha\geq 2)}\sim \epsilon^{2}\ll1.
\end{equation}
Therefore, the growth rate is strongly suppressed if the particle spectrum is shallow, $\alpha< 2$. This suppression is stronger than in the case without the guide field \citep{DemidovLyubarsky2024}.


\section{Tearing instability in the force-free limit}\label{sec:forcefree}

\subsection{The current sheet equilibrium}

In the force-free equilibrium, the plasma pressure is negligible (or can be uniform). Therefore, the current sheet equilibrium is determined only by the magnetic field. The equilibrium assumes that the guide field must be larger inside the current sheet than outside it, such that
\begin{equation}
 B_g(x)=\sqrt{B^2-B_{0y}^2\tanh^2(x/L)},  
\end{equation}
where $B=\text{const}$ is the total magnetic field.

It is important to note that the force-free equilibrium is more realistic than the equilibrium we considered in the previous Section. When a current sheet is formed, the plasma inside is compressed, and the guide field inside the sheet increases due to the "frozen-in" condition until the pressure balance, $B^2/8\pi=\text{const}$, across the sheet is achieved.

Electrons and positrons move in the $y$-$z$ plane at an average velocity $\mathbf{U}_s=U_{sy}\hat{\mathbf{y}}+U_{sz}\hat{\mathbf{z}}$, where $s$ denotes the sort of particles. It is important to note that the particle drift is negligible in the force-free case, $P\ll B^2/8\pi$. The velocity $U$ is simply due to the average particle motion along the magnetic field lines (see Appendix B). The equilibrium current density $\mathbf{j}$ satisfies to Ampère's law
\begin{equation}
    \mathbf{j}=\frac{c}{4\pi}\nabla\times\mathbf{B}.
\end{equation}
Therefore, we have the following estimation for the characteristic values of the macroscopic velocity components:
\begin{equation}\label{driftU}
  \frac{|U_{z}|}{c}\sim \frac{|U_{y}|}{c}\sim \frac{1}{8\pi} \frac{B_{0y}}{e n_0 L}.
\end{equation}
Hence, this velocity is determined only by the magnetic field component $B_{y}$. 

\subsection{The growth rate}\label{subsec:rate1}

The same expressions for the tearing parameter~(\ref{Hdelta}) and the current density perturbation~(\ref{current}) can be used. Therefore, the relationship~(\ref{inc}) remains valid. However, in the force-free limit, the plasma pressure does not affect the distribution of the magnetic field, therefore, we do not have equations that can relate the magnetic field and the average particle energy $\langle\mathcal{E}\rangle$. As a result, the growth rate can be rewritten in the following form
\begin{equation}\label{inc1}
    \gamma(q)\sim \frac{q c}{L}d_e^2 \Delta'\frac{B_{0y}}{B_g}\Big\langle\frac{m c^2}{\mathcal{E}}\Big\rangle^{\!-1},
\end{equation}
where $d_e=c/\omega_{pe}$ is the non-relativistic electron-positron plasma skin depth and $\omega_{pe}=\sqrt{8\pi n_0 e^2/m}$ is the non-relativistic plasma frequency.
The result obtained almost completely coincides with the non-relativistic case (e.g., \citealt{Liu2013}); the only difference is that the thermal velocity of particles is equal to the speed of light $c$, and the particles have a relativistic mass $\Gamma m$.

As in the previous case with the uniform guide field, the maximum growth rate corresponds to $q\sim 1/(2L)$, therefore,
\begin{equation}\label{max1}
    \gamma_{\text{max}}\sim \frac{c\,d_e^2}{L^3}\frac{B_{0y}}{B_g}\Big\langle\frac{m c^2}{\mathcal{E}}\Big\rangle^{\!-1}.
\end{equation}

\subsection{Dependence on the power-law index in the force-free limit}

\begin{figure}
  \centering
  \includegraphics[width=\columnwidth]{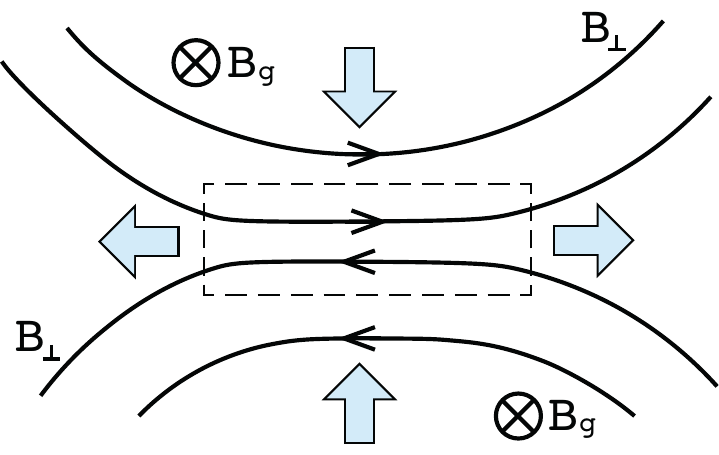}
  \caption{Real current sheets have a finite length. This current sheet can be formed by two current filaments that attract each other due to the Ampère force. The thickness of the current sheet decreases until the incoming magnetic flux begins to dissipate due to reconnection or charge starvation}
\label{f4}
\end{figure}

In nature, there are no infinitely long current sheets, they have a finite length (see Fig.~\ref{f4}). This current sheet can be formed by two flux ropes that carry the current in the same direction. These ropes are attracted to each other by the Ampère force. 
The current sheet between them is compressed until the pressure balance $B^2=\text{const}$ across the sheet is established in the central region due to the compression of the guide field. However, the magnetic pressure is lower on the sides, so as the current sheet thins, the plasma will be squeezed out of it into neighboring regions, which leads to a decrease in the guide field in the central region and further thinning of the current sheet. As long as the current sheet thickness $L$ is much larger than plasma microscopic scales, the tearing instability is too slow and there is no reconnection. The thinning of the current sheet continues until the incoming magnetic flux begins to dissipate. In the case under consideration, dissipation can occur in two ways: tearing instability that leads to magnetic reconnection or charge starvation.

In the case of tearing instability, the magnetic field reconnects, and plasma is ejected to the sides or accumulates in magnetic islands, which leads to a steady state.
However, the maximum current density that can be supported by a plasma is $j_{\text{max}}=2n_0 e c$, which formally means the drift velocity is $U_s\sim c$.
Therefore, according to~(\ref{driftU}), for any ratio $B_{0y}/B_g$ we always can write
\begin{equation}
    L_{\text{min}}\sim \frac{B_{0y}}{8\pi e n_0}.
\end{equation}
At smaller current sheet thicknesses, i.e. when $L<L_\text{min}$, there are not enough charge carriers to provide the required current density and ensure increasingly sharp change in the magnetic field through the current sheet. Therefore, according to Maxwell's equations an electric field $\delta \mathbf{E}_c$ should be generated parallel to $\mathbf{B}$ that heats plasma and leads to the dissipation of magnetic energy with the dissipation rate $W=\delta \mathbf{E}_c\cdot \mathbf{j}$ (however, unlike the tearing mode, there are no "stipes" of the electric field polarity).

At $L=L_{\text{min}}$ the tearing mode rate is
\begin{equation}
    v_\perp\sim \gamma_\text{max}L_\text{min}\sim c\frac{B_g}{\sigma B_{0y}}\left[\langle\mathcal{E}\rangle\langle\mathcal{E}^{-1}\rangle\right]^{-1},
\end{equation}
where $\sigma = B_g^2/(8\pi n_0 \langle \Gamma\rangle mc^2)$ is the magnetization parameter with respect to the guide field. The dependence on the magnetic field strength is clear: for large $\sigma$, the layer thickness $L_{\text{min}}$ increases, leading to a lower tearing rate. Specifically, when $B_g\sim B_{0y}$ and $\sigma\sim 10$, we obtain $v_\perp\sim 0.1c$. 

One can see that the minimum layer thickness has the same value both for $\alpha<2$ and $\alpha\geq 2$. Therefore, in the force-free limit, the tearing instability growth rate does not depend on the particle spectrum and is determined by particles with the lowest energy that mostly participate in the instability.




\section{Relativistic MHD turbulence} \label{sec:turbulence}

\subsection{General remarks\label{subsec:remarks}}

By relativistic MHD turbulence, we understand the situation when the magnetic energy density is significantly greater than the rest energy of the plasma, such that the magnetization parameter is
\begin{equation}
    \sigma=\frac{B^2}{8\pi n\langle \Gamma\rangle mc^2}\gg1.
\end{equation}
Therefore, one can use the force-free limit and neglect the plasma inertia.
Considered pair plasma is relativistic, but all bulk motions are assumed to be non-relativistic or mildly-relativistic.



\begin{figure}
  \centering
  \includegraphics[width=\columnwidth]{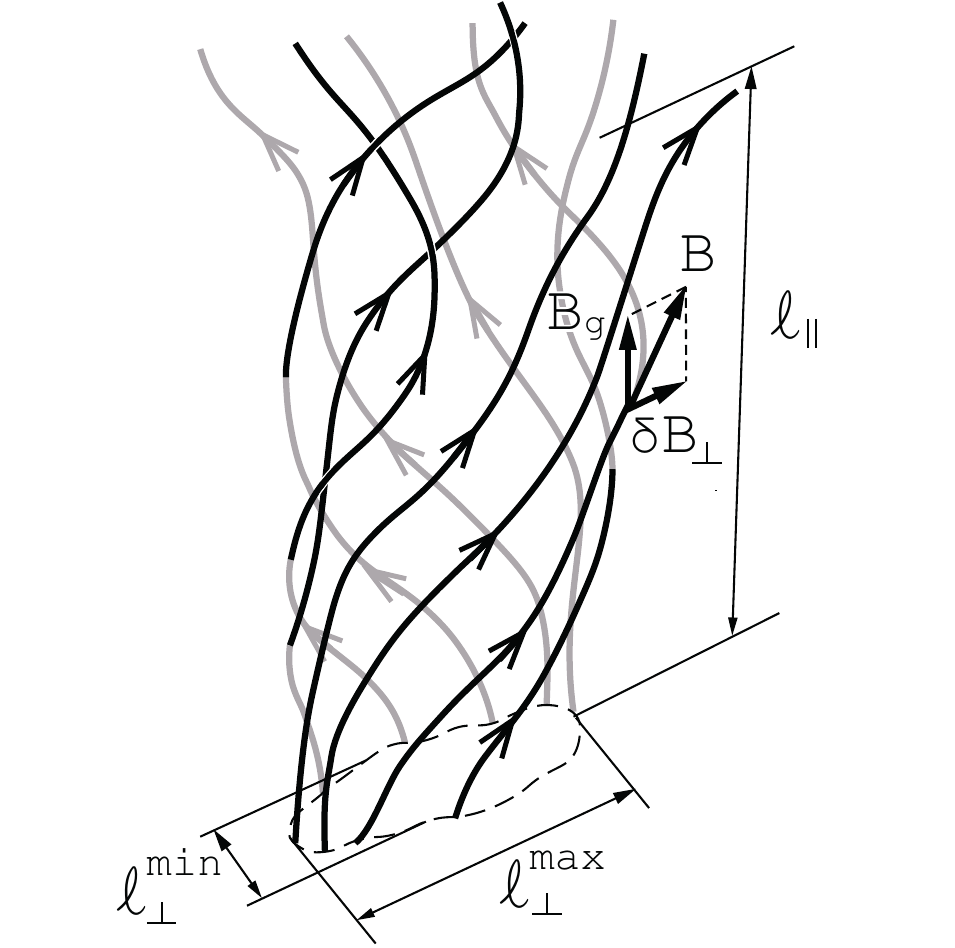}
  \caption{Magnetic field lines within a typical turbulent eddy. The size of the eddy is limited by the region of correlation of magnetic field lines at the considered perpendicular scale. 
  }
\label{f3}
\end{figure}

In general, we cannot assume that 
MHD turbulence is incompressible as in the non-relativistic case;
even if MHD turbulence is driven only by Alfvén modes, during the cascade process fast magnetosonic waves are naturally produced via three-wave interaction (\citealt{ThompsonBlaes1998}, \citealt{Lyubarsky2019}, \citealt{TenBarge2021}). However, the amount of fast modes generated from the Alfvénic turbulence is small. This result is confirmed by numerical simulations both for decaying (\citealt{Cho2005}) and driven turbulence (\citealt{Cho2014}). Moreover, we intend to study the influence of the tearing mode on the turbulent spectrum that occurs 
close to the dissipative scale. In this case 
fast and Alfvén modes are well separated and statistical properties of relativistic turbulence remain the same as in non-relativistic one \citep{TenBarge2021}. Therefore, we can consider only Alfvénic turbulence.
We assume a critically balanced turbulent cascade \citep{GoldreichSridhar1995} when Alfvén waves distort their shape not in many collisions but practically in one. 
It develops mostly in the plane perpendicular to the guide field such that the energy spectrum in the inertial interval is $E(k_\perp)\sim k_\perp^{-5/3}$ and eddies are strongly elongated along the guide magnetic field, which implies two-dimensional anisotropy of MHD turbulence. \cite{Boldyrev2005} suggested that due to dynamical alignment, eddies are anisotropic also in a plane perpendicular to the guide field such that MHD turbulence is essentially three-dimensional. At smaller scales this anisotropy becomes stronger, therefore, eddies on sufficiently small scales turn into a set of current sheets that may become unstable to the tearing instability (\citealt{Loureiro2017}, \citealt{Mallet2017}). Let denote $l_\perp^{\text{min}}$ and $l_\perp^{\text{max}}$ as the minimum and maximum average transverse scales of the eddies (see Fig.~\ref{f3}). According to \cite{Boldyrev2006}, one can write 
\begin{equation}\label{angle}
\theta _l\sim l_\perp^{\text{min}}/l_\perp^{\text{max}}\sim (l_\perp^{\text{min}}/R_\perp)^{1/4},
\end{equation}
where $R_\perp$ is the outer scale perpendicular to the guide field at which turbulence is assumed to be critically balanced and $\theta_l\ll1$ is the dynamic alignment angle. Within this small scale-dependent angle, fluctuations of the plasma velocity $\delta \mathbf{v}_\perp$ and the magnetic field $\delta \mathbf{B}_\perp$ become spontaneously aligned\footnote{It should be emphasized that approach \cite{Chandran2015} considers the dynamic alignment of Elsässer variables, which in the relativistic case $\sigma\gg 1$ are reduced to $\delta\mathbf{z}_\pm=\delta\mathbf{v}\pm c(\delta \mathbf{B}_\perp/B_0)$}. This alignment suppresses the non-linear interaction of the counterpropagating fluctuations at smaller scales such that the cascade time is increased and the energy spectrum becomes slightly flatter $E(k_\perp)\sim k_\perp^{-3/2}$. The dynamical alignment was found in high-resolution relativistic MHD simulations of magnetized decaying turbulence \citep{Chernoglazov2021}, and the results perfectly match with the scaling law~(\ref{angle}). Even though the Boldyrev spectrum $k_\perp^{-3/2}$ is not observed in 3D PIC simulations (e.g. \citealt{Nattila2021}), this might be due to the fact that in such simulations the computational box sizes are usually too small (there is not enough scale separation for a developed cascade and at the outer scale $\theta_l\sim 1$, such that there is no reduction in nonlinear interaction of Alfvén waves).


Alfvén waves propagating along magnetic field lines are responsible for their wandering, which thus dictates the relationship between the transverse and parallel size of eddies. In the critically balanced turbulence, the collision time of counter-propagating Alfvén waves $\tau_c\sim l_\parallel/c$ equals the wave distortion time $\tau_d\sim l_\perp^{\text{max}}/\delta v_{\perp}$. Taking into account that the velocity of turbulent motion at considered scale is $\delta v_\perp\sim c(\delta B_\perp/B_g)$, we obtain
\begin{equation}\label{crit1}
    l^{\text{max}}_\perp B_g\sim l_\parallel \delta B_\perp.
\end{equation}
At the outer scale, we have $\delta B_{\perp0}\sim (R_\perp/R_\parallel)B_g\lesssim B_g$, where $R_\parallel\gtrsim R_\perp$ is the outer scale parallel to the guide field. It also assumes that $\delta v_\perp< c$ at $l_\perp<R_\perp$ and plasma bulk motions are non-relativistic at small scales. 


Taking into account~(\ref{angle}), one can write
\begin{equation}\label{magn}
    l^{\text{max}}_\perp\sim R_\perp\left(l_\perp^{\text{min}}/R_\perp\right)^{3/4}.
\end{equation}
In the inertial range, the energy flux $\delta B_\perp^2/\tau_c$ is conserved across the scales that gives the relation between $l_\parallel$ and $l_\perp^{\text{min}}$:
\begin{equation}\label{parallel}
    l_\parallel/R_\parallel\sim \left(l_\perp^{\text{max}}/R_\perp\right)^{2/3}\sim \left(l_\perp^{\text{min}}/R_\perp\right)^{1/2}.
\end{equation}
At distances larger than $l_\parallel$, spatial correlation between magnetic field lines within an eddy with transverse size $l_\perp^{\text{min}}$ disappears due to differential wandering of magnetic field lines. 


In addition to elongated eddies, there are long plasmoid-unstable current sheets in Alfvénic turbulence. They arise when the largest perturbations collapse or undergo shear without breaking down into smaller perturbations (see review \citealt{Schekochihin2022}). These thin and long intermittent current sheets occupy a very small fraction of the plasma volume but provide a significant fraction of the total magnetic-field dissipation (e.g. \citealt{Zhdankin2014}, \citealt{Zhdankin2016}, \citealt{Chernoglazov2021}). 
To account for the intermittency of MHD turbulence, a more advanced theory must be used (e.g. \citealt{Chandran2015}). Above we examined the scaling laws for the most typical fluctuations that
determine the turbulence energy spectrum. 
However, intermittency is characterized by the rare and "most intense" fluctuations with $\delta B_\perp\sim \delta B_{\perp0}$. 
According to \cite{Mallet2017}, for these fluctuations we have
\begin{equation}\label{int}
    l_\parallel/R_\parallel\sim l_\perp^{\text{max}}/R_\perp\sim (l_\perp^{\text{min}}/R_\perp)^{1/2}.
\end{equation}
Therefore, the length of these sheet-like structures is of the same order as width, such that the duration $\tau_d$ of them is proportional to their length (see also \citealt{Zhdankin2016}). This relation is useful to rewrite in the form $l_\perp^{\text{max}}\sim (\delta B_{\perp0}/B_g)l_\parallel$ that illustrates the dependence on the magnetic field, in particular, one can see that the strong guide field reduces the width of these sheets. The condition $\delta B_\perp\sim \delta B_{\perp0}$ gives rise to $\tau_d\sim \tau_c$ for intemittent structures, therefore, they are also critically balanced.

\subsection{Conditions for tearing-mediated turbulence}

In this section, we derive expressions for transverse sizes of elongated eddies and intermittent current sheets at which they become unstable with respect to tearing mode. Firstly, let us focus on elongated eddies and not on the intermittent structures. 
For simplicity, we use the notation $l_\perp$ instead of $l_\perp^{\text{min}}$.
In the case of tearing-mediated turbulence, the energy cascade is governed by the balance between the linear tearing time and the cascade time $\tau_d\sim l_\perp^{\text{max}}/\delta v_\perp$ (\citealt{Loureiro2017}). It means that
\begin{equation}\label{nl}
    \gamma_{\text{max}}\sim \frac{c}{l_\perp^{\text{max}}}\left(\frac{\delta B_\perp}{B_g}\right),
\end{equation}
where $\gamma_{\text{max}}$ is the maximal growth rate of the linear stage of the tearing instability. This relation also gives some critical value of aspect ratio $l^{\text{max}}_\perp/l_\perp$
above which stable current sheets cannot exist since the linear tearing time exceeds the "lifetime" of the turbulent eddy. 

Therefore, taking into account~(\ref{max1}), one can estimate
\begin{equation}\label{max2}
    \gamma_{\text{max}}\sim \frac{c\, d_e^2}{l_\perp^3}\frac{\delta B_\perp}{B_g}\Big\langle\frac{m c^2}{\mathcal{E}}\Big\rangle^{-1},
\end{equation}
where we assumed that the characteristic current sheet thickness is $L\sim l_\perp$, the guide field is $B_g\sim B_g$ and the reconnecting field is $B_{0y}\sim \delta B_\perp$.

The relationship~(\ref{nl}) can be used to obtain the disruption scale $l_{\text{d}}$ to the tearing-mediated range:
\begin{equation}\label{disrup1}
    \frac{l_{\text{d}}}{R_\perp}\sim \left(\frac{d_e}{R_\perp}\right)^{8/9}\!\!\Big\langle\frac{m c^2}{\mathcal{E}}\Big\rangle^{-4/9},
\end{equation}
If the average inverse Lorentz factor $\langle\Gamma^{-1}\rangle=\langle mc^2/\mathcal{E}\rangle\sim 1$, i.e. $\mathcal{E}_\text{min}\sim mc^2$, then we reproduce the result in the case of magnetized collisionless nonrelativistic pair plasma \citep{Loureiro2018}. 

It is necessary to check whether tearing instability is actually possible in the interval $d<l_\perp\ll l_{\text{d}}$, where $d$ is the lower boundary of the inertial interval
that is determined by the condition that Alfvén waves can be considered in the MHD approximation. In the relativistic case, by using two-fluid MHD equations for pair plasma one can obtain  $d=\sqrt{w/8\pi n_0^2 e^2}$, where $w$ is the enthalpy density \citep{Vega2022a,Vega2024}. In ultrarelativistic case $w\approx(8/3)n_0\langle\mathcal{E}\rangle$, therefore, the dissipative scale can be rewritten as $d\approx d_e\langle\mathcal{E}/mc^2\rangle^{1/2}$, where we neglected the factor $\sqrt{8/3}\sim 1.6$. In strong magnetic fields, when particles are subjected to strong synchrotron cooling, the distribution function is effectively one-dimensional, and the dissipative scale is determined by the same formula, with the only difference that $\mathcal{E}$ should be understood as the longitudinal energy and averaging is performed over the longitudinal momentum (see Appendix C).  

If the magnetic field is not frozen-in everywhere, it does not create tension in the plasma that is the necessary ingredient for magnetic reconnection. 
In the case of a hard particle spectrum $\alpha<2$, at the dissipative scale $d$, particles with energies $\mathcal{E}\gtrsim\langle\mathcal{E}\rangle$ due to their large inertia are no longer frozen into the plasma 
over the Alfvén wave period $\sim l_\parallel/c$ (which also corresponds to the cascade time, $\tau_c$, due to critical balance), but the lowest-energy particles, $\mathcal{E}\sim \langle\mathcal{E}^{-1}\rangle^{-1}$, remain frozen-in. It can be assumed that the tearing instability operates down to the scale 
$d_e \langle mc^2/\mathcal{E}\rangle^{-1/2}$ with the same growth rate 
$\gamma_{\text{max}}$, as the dominant contribution comes from low-energy 
particles. However, Landau damping on high-energy particles must be considered 
at scales $l_\perp < d$. This damping causes turbulence to decay, leading to 
an increasing deviation from the $k_\perp^{-3/2}$ spectrum as $k_\perp$ rises. 
Nevertheless, a small number of high-energy particles and their large relativistic mass $\sim \Gamma^3 m$ can reduce the damping, 
allowing the tearing mode of the low-energy component of the plasma to remain dominant.
The interplay between the tearing mode and Landau damping depends on the exact form of the distribution function and values of $\alpha$ and $\epsilon$. The detailed consideration of this  requires further study beyond the scope of this paper. 

If Landau damping of high-energy particles is important, then we should use the dissipative scale $d$, therefore, one can rewrite~(\ref{disrup1}) in the following form
\begin{equation}\label{disrup2}
    \frac{l_{\text{d}}}{d}=\left(\frac{R_\perp}{d}\right)^{1/9}\!\!\!\left[\langle\mathcal{E}\rangle\langle\mathcal{E}^{-1}\rangle\right]^{-4/9},
\end{equation}
where the last factor with average energies plays a role only when $\alpha<2$; otherwise, it is of the order of unity. Therefore, at $\alpha<2$ the inequality $l_{\text{d}}>d$ is fulfilled when the ratio of the outer scale $R_\perp$ to the dissipative scale $d$ is larger than $[\langle\mathcal{E}\rangle\langle\mathcal{E}^{-1}\rangle]^{4}$, otherwise the tearing instability is not possible in the inertial interval $l_\perp>d$. Indeed, according to~(\ref{disrup1}) for small-scale separation, the disruption scale is very close to the plasma skin-depth of low-energy particles $d_e\langle mc^2/\mathcal{E}\rangle ^{-1/2}$, which is smaller than the dissipative scale $d$.  

In the case of intermittent current sheets, we have the same equations~(\ref{nl}) and~(\ref{max2}), but for $l_\perp^{\text{max}}$ the scaling law~(\ref{int}) should be used. As a result, we obtain
\begin{equation}
    \frac{l'_{\text{d}}}{R_\perp}\sim \left(\frac{d_e}{R_\perp}\right)^{4/5}\!\!\Big\langle\frac{m c^2}{\mathcal{E}}\Big\rangle^{-2/5} 
\end{equation}
and
\begin{equation}
    \frac{l'_\text{d}}{d}=\left(\frac{R_\perp}{d}\right)^{1/5}\!\!\!\left[\langle\mathcal{E}\rangle\langle\mathcal{E}^{-1}\rangle\right]^{-2/5}.
\end{equation}
Therefore, size of the system and particle spectrum have a weaker effect on the disruption of an intermittent current sheet: for $l'_{\text{d}}>d$ at $\alpha<2$, we obtain $R_\perp/d>[\langle\mathcal{E}\rangle\langle\mathcal{E}^{-1}\rangle]^{2}$.

If Landau damping of high-energy particles can be neglected at $d_e\langle mc^2/\mathcal{E}\rangle^{-1/2}<l_\perp<d$ such that the real dissipative scale of the turbulence is $d_{\text{min}}=d_e\langle mc^2/\mathcal{E}\rangle^{-1/2}$, then disruption scales are
\begin{equation}
    \frac{l_\text{d}}{d_{\text{min}}}=\left(\frac{R_\perp}{d_{\text{min}}}\right)^{1/9}, \quad \frac{l'_\text{d}}{d_{\text{min}}}=\left(\frac{R_\perp}{d_{\text{min}}}\right)^{1/5}.
\end{equation}
In this case, there is no dependence on the particle spectrum, and high-energy particles do not play a role, such that $l_{\text{d}}>d_{\text{min}}$ always holds true.

Despite the fact that the thickness of the intermittent current sheets is very small, these structures themselves are not formed as a result of a cascade (\citealt{Chandran2015}) and can be considered as discontinuities of the outer amplitude $\delta B_\perp\sim \delta B_{\perp,0}$, therefore, the energy disappears from the system at the outer scale and turbulence spectrum is not modified. Unstable elongated eddies modify the turbulence spectrum at $d<l_\perp<l_d$. Obviously, this modification should be the same as in the nonrelativistic
case, i.e. $E(k_\perp)\sim k_\perp^{-3}$ (\citealt{Loureiro2018}).  

One can see that at the ratio $R_\perp/d<10^9$ the disruption scale of elongated eddies $l_{\text{d}}$ is of the same order as the dissipative scale $d$.
In order to clearly see tearing-mediated turbulence with the modified spectrum, the disruption scale must be at least an order of magnitude greater than $d$. This is possible only if $R_\perp/d\geq 10^9$. For intermittent current sheets one needs smaller ratio $R_\perp/d\geq 10^5$. Such scale ratios can be encountered in real astrophysical systems (see Section 5), but cannot be obtained in modern PIC and MHD simulations. However, numerical simulations of turbulent collisionless plasma are affected by numerical noise: in MHD simulations, this arises from the finite cell size, while in PIC simulations, it is due to the limited number of particles per cell. These effects either generate numerical resistivity or introduce small initial perturbations in the current sheets, enhancing the tearing mode's growth rate and increasing the disruption scale (this is mostly relevant for intermittent current sheets). 

\subsection{Charge starvation}

However, even if we have such a system with $l_{\text{d}}>d$ and current sheets are sufficiently elongated such that tearing instability is possible in the inertial range, we also have to check whether there are enough charge carriers to support the forming current layers. This problem does not exist in the non-relativistic case, but it can arise in relativistic MHD turbulence. Let us compare the maximum current density that can be supported by plasma,
\begin{equation}
    j_{\text{max}}=2n_0 e c,
\end{equation}
with the electric current in 
a current sheet (see also \citealt{ChenYuan2022}, \citealt{Natilla2022}) 
\begin{equation}
\delta j_l\sim \frac{c}{4\pi}\frac{\delta B_\perp}{l_\perp}.
\end{equation}
If the maximum current density $j_{\text{max}}$ is less than $\delta j_l$ at $l_\perp = l_{\text{d}}$,  then there are not enough charge carriers to sustain the current sheet. Consequently, the turbulence decays before the tearing mode comes into play.
Taking into account that $ d=\sqrt{\sigma}r_g$, where $\sigma= B_g^2/(8\pi n_0\langle\Gamma\rangle mc^2)$ and $r_{g}=\langle\mathcal{E}\rangle/eB_g$ is the average Larmor radius in the guide field, the inequality $\delta j_l>j_{\text{max}}$ for elongated eddies can be rewritten in terms of the magnetization parameter:
\begin{equation}\label{sigma}
    \sigma>\sigma_{\text{cr}}=
        \left(\frac{R_\parallel}{R_\perp}\right)^2\!\left(\dfrac{R_\perp}{d}\right)^{2/3}\!\!\left[\langle\mathcal{E}\rangle\langle\mathcal{E}^{-1}\rangle\right]^{-2/3},
\end{equation}
where $\sigma_{\text{cr}}$ is the critical magnetization parameter above which charge starvation is occurred. This relation can be rewritten in terms of $\Tilde{\sigma}=(R_\perp/R_\parallel)^2\sigma$, where $\Tilde{\sigma}$ is the magnetization parameter with respect to the outer magnetic field perturbation $\delta B_{\perp0}$.

For intermittent structures the charge starvation condition $\delta j_l>j_{\text{max}}$ at $l_\perp=l'_{\text{d}}$ can be rewritten as
\begin{equation}\label{sigma1}
\sigma'>\sigma'_{\text{cr}}=\left(\frac{R_\parallel}{R_\perp}\right)^2\!\left(\frac{R_\perp}{d}\right)^{2/5}\!\!\left[\langle\mathcal{E}\rangle\langle\mathcal{E}^{-1}\rangle\right]^{-2/5}.
\end{equation}
One can see that if the particle spectrum is not important, i.e. $\alpha\geq 2$, then $\sigma'_{\text{cr}}<\sigma_{\text{cr}}$. In other words, more charge carriers are needed for intermittent current sheets than for elongated eddies, since the field jump across the intermittent sheet is large $\sim \delta B_{\perp0}$, but the layer thickness of both structures is comparable.

If Landau damping of high-energy particles is negligible for $\alpha<2$ the critical magnetization parameter formulas should be modified by substituting $d\rightarrow d_{\text{min}}$ and removing the factor $\langle\mathcal{E}\rangle\langle\mathcal{E}^{-1}\rangle$.

\cite{Natilla2022} examined the charge starvation condition for the Maxweillian plasma at the dissipative scale $l_\perp\sim d<l_{\text{d}}$ thereby slightly softening the condition on the critical magnetization. 
They conducted two simulations with low and high magnetization parameters, demonstrating that in the latter case, Alfvén waves indeed enter the charge starvation regime. 

\section{Astrophysical applications}

The spectral index $\alpha$ varies depending on the astrophysical context and the processes involved, but the typical value is $\alpha>2$. The harder particle spectrum with $\alpha\rightarrow 1$ can be seen in regions of magnetic reconnection in highly magnetized plasma (e.g. \citealt{Guo2014},\citealt{Werner2016}). Moreover, the higher the plasma magnetization parameter $\sigma$, the closer the spectral index $\alpha$ is to unity. However, more recent simulations (\citealt{Petropoulou2018}, \citealt{French2023}) show convergence towards $\alpha\sim 2$ at late times of the simulation, but with the same dependence $\alpha$ on $\sigma$ (see also review \citealt{Guo2024}). It should be noted that in the above studies the particle spectra with $\alpha<2$ were obtained when the guide field was weak with respect to magnetic field perturbations. Large guide fields lead to softer spectra and lower maximum energies (\citealt{French2023}). 

\begin{figure}
  \centering
  \includegraphics[width=\columnwidth]{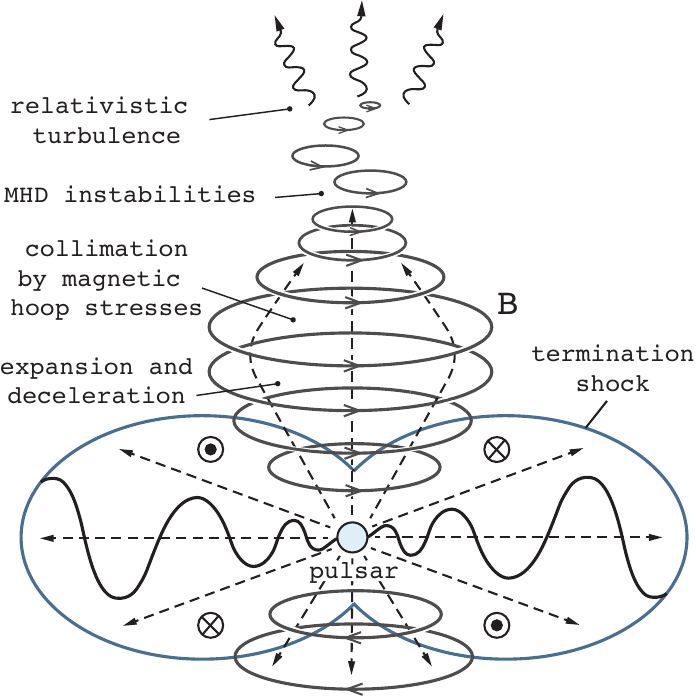}
  \caption{The magnetized pulsar wind at high latitudes is collimated by magnetic hoop stresses. MHD instabilities and shrinking of magnetic loops give rise to relativistic turbulence
  }
\label{f6}
\end{figure}

These results can be applied to pulsar wind nebulae (PWNe) where the radio spectrum for most of them is flat that implies $\alpha\rightarrow 1$ (\citealt{Reynolds2001},\citealt{Green2014}). \cite{Lyutikov2019}, \cite{Luo2020} considered simple 1D MHD and radiation model, where the pulsar wind just beyond the termination shock still has relatively large magnetization, 
but is highly turbulent such that the turbulence dominates the flow. This turbulence self-consistently generates reconnecting current sheets within the bulk of the Nebula, leading to magnetic field dissipation and strong particle acceleration with $\alpha<2$ (there is also acceleration on the termination shock, but it gives $\alpha>2$). However, if the wind is highly magnetized at the termination shock, the shock should be weak, which means that postshock flow remains relativistic and radial without any turbulence. High magnetization flow can exist near the axis of the PWNe, where magnetization is high because the magnetic field does not change the sign in the high-latitude part of the pulsar wind, preserving energy until dissipation beyond the termination shock (\citealt{Lyubarsky2012}). This shock is weak and close to the pulsar at high latitudes. Beyond the shock, the flow expands and decelerates, which eventually leads to collimation due to magnetic hoop stresses, which are not balanced by either the poloidal magnetic field or the plasma pressure. This converging flow is unstable (\citealt{Sobacchi2018}) since even small perturbations due to kink (\citealt{Begelman1998}, \citealt{Mizuno2011}) or Kelvin-Helmholtz (\citealt{Begelman1999}) instabilities destroy it so that magnetic loops come apart and then shrink independently of each other producing relativistic turbulence (see Fig.~\ref{f6}). This small region of turbulent plasma may be responsible for gamma flares (\citealt{Tavani2011}, \citealt{Abdo2011}). The April 2011 flare clearly shows an additional synchrotron component above the quiescent emission. The time-averaged spectrum is hard, indicating the power-law particle distribution with the slope $\alpha=1.6$ \citep{Buehler2012}. 

Let us estimate scale separation $R_\perp/d$ for this turbulent plasma. Since there are many uncertainties in the parameters of the problem, we provide only very rough estimations.
The evolution of the outflow occurs in a symmetrical manner: it is cooled when it expands and then converges and accelerates. The flow is maximally expanded at $r_m\sim \theta_0^2 R_{TS}$ (\citealt{Lyubarsky2012}), where $\theta_0=\pi/2 - \alpha$ is a polar angle, exceeding which the striped wind zone is formed, $\alpha$ is the pulsar inclination angle and $R_{TS}$ is the equatorial radius of the termination shock. The converging point is $r_c\sim 2r_m$. However, at the convergence phase, the outflow is unstable, and turbulence occurs. We assume that the radius of the outflow decreases by approximately an order of magnitude, from which it follows that the number density increases by approximately two orders of magnitude, i.e. $n_c\sim 10^2n_m$.

Since the termination shock at high latitudes is weak and plasma enters the shock highly obliquely, one can assume the radial flow during the expansion phase. Therefore, the number density of plasma at $r\sim r_m$ can be estimated as 
\begin{equation}
    n_m\sim \mathcal{M}\frac{\Omega B_{LC}}{2\pi ce}\left(\frac{R_{LC}}{\theta_0^2R_{TS}}\right)^2,
\end{equation}
where $\mathcal{M}\sim 10^6$ is the pair plasma multiplicity, $\Omega$ is the pulsar angular velocity, $B_{LC}=B_*(R_*/R_{LC})^3$ is the magnetic field at the light cylinder $R_{LC}\sim c/\Omega$. In turn, $R_{TS}$ can be found by balancing the wind pressure $p_{\text{wind}}\sim L_{\text{sd}}/4\pi r^2 c$, where $L_{\text{sd}}=B_*^2 R_*^6\Omega^4/c^3$ is the spin-down luminosity of the pulsar, with the ambient pressure $p_{\text{ext}}\sim 10^{-7}$ dyn/cm$^2$ (this pressure corresponds to a nebula expansion with velocity $v\sim 2000$ km/s within a medium with the density $4\times 10^{-24}$ g/cm$^3$). 
Combining these data and taking into account that $n_c\sim 10^2 n_m$ we obtain
\begin{equation}
    n_c\sim 10\mathcal{M}_6p_{\text{ext},-7}\left(\frac{P}{1\,\text{s}}\right)^2\!\left(\frac{\theta_0}{\pi/6}\right)^{\!-4} \,\, \text{cm}^{-3},
\end{equation}
where we put $B_*=10^{12}$ G and $R_*=10^6$ cm. In particular, for Crab $P=0.033$ s, therefore, $n_c\sim 10^{-2}$ cm$^{-3}$.

Since the plasma enters this area already cold $\langle\tilde{\mathcal{E}}\rangle\sim mc^2$, the average magnetization parameter in the co-moving frame is $\sigma={\tilde B_c}^2/(4\pi \tilde n_c mc^2)$. At large distances only toroidal component of the magnetic field survives, and at small $\theta_0$ it is perpendicular to the wind velocity, therefore, $\tilde B_c=B_c/\Gamma_{\text{wind}}$ and $\tilde n_c=n_c/\Gamma_{\text{wind}}$. It gives $\sigma=B_c^2/(4\pi mc^2n_c\Gamma_{\text{wind}})$, where all the quantities are measured in the lab frame, where the pulsar is at rest. 
The wind is magnetized with $\sigma \sim 100$ after the termination shock but before the dissipation (e.g. \citealt{ZrakeArons2017}). Therefore, assuming that the Lorentz factor of the wind is $\Gamma_{\text{wind}}\sim 10$, the magnetic field is 
\begin{equation}
   \!\!\! B_c\!\sim\! 10^{-1}\sigma_{100}^{1/2}\Gamma^{1/2}_{\text{wind},1}\mathcal{M}^{1/2}_6p_{\text{ext},-7}^{1/2}\!\! \left(\frac{P}{1\,\text{s}}\right)\!\!\left(\frac{\theta_0}{\pi/6}\right)^{\!\!-2}\!\! \text{G}.
\end{equation} 
The outer scale of the turbulence can be estimated as $R_\perp\sim 0.1\theta_0 r_m\sim0.1\theta_0^3R_{TS}$ (\citealt{Lyubarsky2012}). The plasma skin-depth in the co-moving frame is $d\sim \sqrt{mc^2/4\pi \tilde n_c e^2}$. Therefore, 
\begin{equation}
    \frac{R_\perp}{d}\sim 10^6\mathcal{M}_6^{1/2}\Gamma_{\text{wind},1}^{-1/2}\left(\frac{P}{1 \,\text{s}}\right)^{-1}\!\!\left(\frac{\theta_0}{\pi/6}\right).
\end{equation}
For Crab pulsar $R_\perp/d\sim 10^7$, corresponding disruption scales are $l_d\sim 5d$ for elongated eddies and $l'_d\sim25 d$ for intermittent current sheets.

The corresponding magnetization parameters at which charge starvation occurs are 
\begin{equation}
    \sigma_{\text{cr}}\sim 10^4 \mathcal{M}_6^{1/3}\Gamma_{\text{wind},1}^{-1/3}\left(\frac{P}{1 \,\text{s}}\right)^{\!\!-2/3}\!\!\left(\frac{\theta_0}{\pi/6}\right)^{\!\!2/3}
\end{equation}
for elongated eddies and
\begin{equation}\label{cr1}
    \sigma'_{\text{cr}}\sim 10^2 \mathcal{M}_6^{1/5}\Gamma_{\text{wind},1}^{-1/5}\left(\frac{P}{1 \,\text{s}}\right)^{\!\!-2/5}\!\!\left(\frac{\theta_0}{\pi/6}\right)^{\!\!2/5}
\end{equation}
for intermittent current sheets. Since $\sigma\sim 10^{2}$, the above rough estimations are not sufficient to determine whether charge starvation occurs in large-scale intermittent current sheets. Nevertheless, in any case, this process also leads to dissipation and particle heating along the guide field. When the magnetic energy begins to dissipate, due to acceleration and heating of the plasma, the scale ratio $R_\perp/d$ decreases. Therefore, one can expect the magnetic field dissipation in different regimes. Assuming $\alpha<2$ (\citealt{Buehler2012}), the factor $[\langle\tilde{\mathcal{E}}\rangle\langle\tilde{\mathcal{E}}^{-1}\rangle]^{-1}$ decreases disruption scales $l_d$ and $l'_d$. This effect is more pronounced for elongated eddies, potentially leading to $l_d<d$. As a result, elongated eddies are not disrupted by tearing instability; instead, Landau damping leads to magnetic field dissipation. With a decrease in $\sigma$, tearing instability of intermittent current sheets might occur instead of charge starvation. However, uncertainties in particle spectra, number density, and the outflow Lorentz factor prevent drawing precise conclusions.

The rapid gamma-ray flares observed in the Crab Nebula, exceeding the synchrotron burn-off limit (e.g., \citealt{Kargaltsev2015}), could be explained by assuming small pitch angles for accelerated particles (see also \citealt{Uzdensky2011}, \citealt{Cerutti2013}). Beams of such particles are naturally produced at X-points in reconnecting current sheets and also can be produced by the induced electric field due to charge starvation. 

The next example is AGN jets. We assume that the energy distribution of non-thermal leptons is described by the broken power-law spectrum with the spectral index for the highest energies $\alpha\geq 2$ \citep{Tavecchio1998}. It means that one can neglect the influence of the particle spectrum and put $\langle\tilde{\mathcal{E}}\rangle\sim \mathcal{E}_{\text{min}}\sim 10^2mc^2$ in the co-moving frame. In Poynting-dominated jets, the magnetic energy gives the main contribution to the luminosity, therefore, $L=(B^2/4\pi)c R^2$. The average Larmor radius of particles is $r_g\sim\langle\tilde{\mathcal{E}}\rangle/e \tilde{B}$, where $\tilde{B}=B/\Gamma_{\text{jet}}$ is the co-moving magnetic field and $\Gamma_{\text{jet}}$ is the Lorentz factor of the jet. Using the relation $d=\sqrt{\sigma}r_g$, we obtain
\begin{equation}\label{Rd}
    \frac{R}{d}\sim\sqrt{\frac{4\pi e^2 L}{c\sigma\Gamma_{\text{jet}}^2\langle\tilde{\mathcal{E}}\rangle^2}}.
\end{equation}
Substituting typical parameters of AGN jets for the critical magnetization above which the charge starvation occurs, we obtain
\begin{equation}
    \sigma_{\text{cr}}\sim 6.2\times 10^{5}L_{45}^{1/4}\left(\frac{\Gamma_{\text{jet}}}{10}\right)^{-1/2}\!\!\!\left(\frac{\langle\tilde{\mathcal{E}}\rangle}{10^2mc^2}\right)^{-1/2},
\end{equation}
where $L=10^{45}L_{45}$ erg/s. For intermittent current sheets $\sigma'_{\text{cr}}\sim 7\times 10^3$. The magnetization parameter in AGN jets 
is much lower (otherwise they were accelerated to much higher Lorentz factors than are observed). Therefore, the charge starvation is not achieved. 

According to~(\ref{Rd}), $R/d\sim 10^{10}>10^9$; therefore, elongated eddies become thin enough and can modify the turbulence spectrum, i.e., tearing-mediated turbulence is possible in AGN jets. Corresponding disruption scales are $l_\text{d}\sim 10d$ (for elongated eddies) and $l'_\text{d}\sim 100d$ (for intermittent current sheets). 

The presence of turbulence and current sheets leads to the fact that the distribution of particles becomes anisotropic. The assumption of an anisotropic distribution of emitting electrons helps resolve the paradox of low radiative efficiency in jets of BL Lac objects (\citealt{Sobacchi2019}, \citealt{Sobacchi2021}, \citealt{Sobacchi2023}). Modeling their multi-wavelength spectra typically assumes isotropic particle distribution, leading to the conclusion that the magnetic energy in the energy release zone is much lower than the energy of the radiating particles, i.e. $\sigma\ll 1$ (e.g., \citealt{Tavecchio2016}). This contradicts the current view that relativistic jets are magnetically dominated and should release most of their energy during equipartition such that $\sigma\sim 1$.

\section{Discussion and conclusions}\label{sec:conclusions}


In this work, we considered the collisionless tearing instability in a relativistic pair plasma with a power-law distribution function with the presence of the guide field. In the case of the uniform guide field, when the current sheet is supported by plasma pressure, the tearing mode is strongly suppressed as the particle spectrum hardens $\alpha<2$. On contrary, at $\alpha\geq2$ the tearing instability gives the same growth rate as in the case of relativistic Maxwellian plasma \citep{Zelenyi1979}. In the force-free limit, when plasma pressure does not play any role, the tearing instability growth rate becomes independent of the particle spectrum. This occurs because the layer thickness cannot be reduced to the scale of the average Larmor radius of the particles. At a larger thickness $L_{\text{min}}\gg r_g$, charge starvation arises in the current sheet, inducing an electric field and dissipation of the magnetic field. Since 
$L_{\text{min}}$ is independent of $\alpha$, the maximum possible growth rate of the tearing mode remains unaffected by the particle spectrum.

We also have found at what conditions the spectrum of relativistic MHD turbulence in pair plasma with power-law distribution function can be modified by the tearing instability. 
Namely, we derived expressions for transverse sizes of small-scale elongated eddies and intermittent current sheets unstable to tearing mode. According to the obtained results, in the case of a sufficiently hard particle spectrum $\alpha<2$, the disruption scale of current sheets may turn out to be smaller than the dissipative scale $d$, which is not observed for the Maxwellian distribution function. However, even though the tearing instability can be suppressed, at the kinetic scale $l_\perp<d$ the Alfvén waves are damped by the Landau mechanism, which also heats the plasma and accelerates the particles along the guide field.

One of the main differences between relativistic MHD turbulence and non-relativistic one is that there might not be enough charge carriers to support the current sheets that arise in the turbulent plasma.
An expression for the critical magnetization parameter above which the tearing mode is suppressed by charge starvation is determined. 


The obtained results can be useful for constructing self-consistent models of particle acceleration in relativistic astrophysical sources.


\begin{acknowledgments}
We acknowledge useful discussions with Sasha Philippov, Dmitry Uzdensky, Joonas Nättilä and Vladimir Zhdankin. This research was supported by the Israel Science Foundation under grants 2067/19 and 2126/22.
\end{acknowledgments}



\appendix

\section{Non-magnetized high energy particles}

If $\alpha<2$ and the guide field satisfies the inequality
\begin{equation}\label{limitsB}
B_{0y}\sqrt{\frac{\mathcal{E}_{\text{min}}}{eB_{0y}L}}<B_g<B_{0y}\sqrt{\frac{\mathcal{E}_{\text{max}}}{eB_{0y}L}},
\end{equation}
high-energy particles do not "feel" the guide field. 
For high-energy particles, one should consider the classical problem of tearing instability, as if the guide field were absent (\citealt{DemidovLyubarsky2024}). For such particles, the effective conductivity can be estimated using the Drude formula with the effective collision time $\tau_\text{coll}\sim 1/qc$, therefore 
\begin{equation}\label{max}
    \sigma_{\text{eff}}\sim \frac{2n_H e^2}{m}\frac{1}{qc}\left(\frac{mc^2}{\mathcal{E}_\text{max}}\right)\delta E_z,
\end{equation}
where $n_H$ is the number density of high-energy particles with $\mathcal{E}\sim\mathcal{E}_{\text{max}}$.
In this situation we have the double tearing layer: the first layer $|x|\lesssim l$, with low-energy particles and the second layer of the size $|x|\lesssim l_{\text{max}}=\sqrt{\mathcal{E}_{\text{max}}L/eB_{0y}}$ with high-energy particles. In the overlapping region $|x|\lesssim \text{min}(l,l_{\text{max}})$ the total current density is
\begin{equation}\label{jz}
    \delta j_z\approx\frac{2n_L e^2}{m}\frac{m c^2}{\mathcal{E}_{\text{min}}}\left(\frac{1}{\gamma}+\frac{n_H}{n_L}\frac{\epsilon}{q c}\right)\delta E_z,
\end{equation}
where $n_L=n_0-n_H\gg n_H$ is the number density of low energy particles and $\epsilon=\mathcal{E}_{\text{min}}/\mathcal{E}_{\text{max}}\ll 1$. 
Therefore, one can see that high-energy particles are  important only in the case $\gamma > cn_L/(\epsilon L n_H)$ (the $\delta A$-constant approximation assumes $q>1/(2L)$). 
However, $v_\perp\sim \gamma_{\text{max}}L\leq c$, therefore, $\gamma_\text{max} \leq c/L$ and the contribution of high-energy particles to $\delta j_z$ can be neglected.

\section{drift velocity}

The equilibrium of the current sheet is determined by the balance between the pressure gradient and the Lorentz force,
\begin{equation}
    -\nabla\cdot \mathbf{P}+\frac{1}{c}[\,\mathbf{j}\times\mathbf{B}]=0,
\end{equation}
where $\mathbf{B}=B_0\tanh(x/L)\hat{\mathbf{y}}+B_g\hat{\mathbf{z}}$ is the total magnetic field, $\mathbf{P}=P_\perp \mathbf{I}+(P_\parallel-P_\perp)\mathbf{b}\otimes\mathbf{b}$
is the pressure tensor, $\mathbf{I}$ is the unit tensor and $\mathbf{b}=\mathbf{B}/B$ is the unit vector along the total magnetic field. In the isotropic case, one can put $P_\perp=P_\parallel=P$.

The current density is determined by the Maxwell equation $\mathbf{j}=(c/4\pi)\nabla\times\mathbf{B}$. Since the reconnecting magnetic field varies only with $x$-coordinate and the guide field is constant, we obtain that $\mathbf{j}$ is directed along $z$-axis. The guide field $\mathbf{B}_g$ does not contribute to the force balance equation since it is directed along the total current density. Therefore, we obtain the same result as for the usual Harris current sheet without the guide field:
\begin{equation}\label{currentA}
    \mathbf{j}=-\frac{1}{B_y}\frac{\text{d}P_\perp}{\text{d}x}\hat{\mathbf{z}}=\frac{cB_{0y}}{4\pi L}\text{sech}^2\left(\frac{x}{L}\right)\hat{\mathbf{z}}.
\end{equation}
Thus, the plasma experiences a total drift only along the $z$-axis. 

However, this result seems paradoxical. With a sufficiently strong guide field, all Larmor circles are now oriented in a plane perpendicular to $\mathbf{B}_g$. Therefore, the magnetization current, as well as $\nabla B$-current, should also flow mostly perpendicular to the guide field, i.e. perpendicular to the $z$-axis. Hence the following question arises: can the plasma really maintain a current \textit{only} along the guide field?

Since the magnetic field changes adiabatically along the particle trajectories, one can use the guiding center approximation. Thus, the total current density can be written in the following form \citep{Northrop1961,Bellan2006}
\begin{equation}
    \mathbf{j}=\sum_s q_s \int\left(\dot{\mathbf{R}}_\perp+\mathbf{b}v_\parallel\right)f_s(\mathbf{p})\text{d}\mathbf{p}+c\,[\nabla\times \mathcal{M}],
\end{equation}
where $\dot{\mathbf{R}}_\perp$ is the drift velocity perpendicular to the magnetic field that is induced by the $\nabla B$-drift, $\mathcal{M}_s$ is the plasma magnetic moment per unit volume that can be written as
\begin{equation}
    \mathcal{M}=-\sum_s\int \mu_s\mathbf{b} f_s(\mathbf{p})\text{d}\mathbf{p}=-\frac{P_\perp\mathbf{b}}{B},
\end{equation}
where $\mu=K_\perp/B$ is the magnetic moment of a Larmor circle.

Taking into account that $\dot{\mathbf{R}}_\perp=(\mu c/q_s B)[\mathbf{b}\times \nabla B]$, the $\nabla B$-current becomes
\begin{equation}
    \sum_s q_s\int \dot{\mathbf{R}}_\perp f_s(\mathbf{p})\text{d}\mathbf{p}=\frac{cP_\perp}{B^2}[\mathbf{b}\times\nabla B].
\end{equation}
This current cancels part of the magnetization current, therefore,
\begin{equation}\label{cur1}
    \mathbf{j}=2e n \langle v_\parallel\rangle \mathbf{b}-\frac{cP_\perp}{B}[\nabla\times\mathbf{b}]+\frac{c}{B}[\mathbf{b}\times\nabla P_\perp]=2e n \langle v_\parallel\rangle \mathbf{b}+\frac{c}{B}\frac{\text{d}(P_\perp B_g)}{\text{d}x}\hat{\mathbf{y}}-\frac{c}{B}\frac{\text{d}(P_\perp b_y)}{\text{d}x}\hat{\mathbf{z}}.
\end{equation}
The current along $y$-axis is absent when the $y$-component of the drift along the magnetic field, $\langle v_\parallel\rangle b_y$, compensates for the $y$-component of the transverse drift, i.e.
\begin{equation}\label{dr}
    \langle v_\parallel\rangle =-\frac{c}{2enB b_y}\frac{\text{d}(P_\perp B_g)}{\text{d}x}.
\end{equation}
One can easily check that if we substitute this velocity into~(\ref{cur1}) it exactly gives the total current~(\ref{currentA}). Therefore, $j_z=2n e U_z$ and $j_y=2n eU_y=0$, where $U_z$ and $U_y$ are total drift velocities along $z$ and $y$ axes respectively. Assuming $n(x)=n_0\text{sech}^2(x/L)$, which is correct for the constant velocity profile, we obtain
\begin{equation}
    \frac{U_z}{c}=\frac{B_0}{8\pi n_0 e L}\sim \frac{\langle\mathcal{E}\rangle}{eB_{0y} L}
\end{equation}
that is used in~(\ref{two}). Therefore, although the formula looks like an ordinary diamagnetic drift along the $z$-axis as in the case of the Harris current sheet with $B_g=0$, it is actually a combination of diamagnetic drift perpendicular to $\mathbf{B}$ and the "fine-tuned" longitudinal drift along $\mathbf{B}$ that is ensured by the correct choice of the distribution function (which in turn adapts to the given geometry of the magnetic field through the Vlasov equation, otherwise such equilibrium would be impossible).

The resulting drift can be taken into account by using the Galilean transform for energy, such that $f_s=f_s(\mathcal{E}-U_s P_z)$. However, near the neutral plane, where we are interested in the form of the distribution function, one can neglect the term $U_s P_z\approx U_s p_z$ in comparison with the energy of ultrarelativistic particles $\mathcal{E}\approx pc$, since $U_s\ll c$.

In the force-free limit, we have $P_\perp\ll B^2/8\pi$, and the pressure may be constant through the current sheet, which means that one can neglect all drifts across the field. Therefore
\begin{equation}
    \mathbf{j}=2en\langle v_\parallel\rangle \mathbf{b}.
\end{equation}
This is in full agreement with the definition of a force-free field $\mathbf{j}\times\mathbf{B}=0$, such that $\mathbf{j}\parallel\mathbf{B}$.

\section{The dispersion equation for Alfvén waves in highly magnetized plasma}


Assuming harmonic excitations, Maxwell equations can be reduced to the following form
\begin{equation}\label{electric}
[\mathbf{k}\times[\mathbf{k}\times\mathbf{E}]]+\left(\frac{\omega}{c}\right)^2\!\mathbf{E}=-\frac{4\pi\text{i}\omega}{c^2}\mathbf{j},
\end{equation}
In the force-free field, the total current is directed along $\mathbf{B}$. Let us assume that $\mathbf{B}\parallel\hat{\mathbf{z}}$ and $\mathbf{k}=(k_\perp,0,k_\parallel)$. Then the current density can be expressed via the electric field by using Ohm's law $j_z=\sigma E_z$, where $\sigma=\sigma(\omega,\mathbf{k})$ is the electric conductivity.

When $\mathbf{E}$ is perpendicular to the $\mathbf{k}$-$\mathbf{B}$ plane we obtain the dispersion equation $\omega=kc$. It corresponds to the extraordinary mode (E-mode). When $\mathbf{E}$ lies within the $\mathbf{k}$-$\mathbf{B}$ plane the wave is called the ordinary mode (O-mode). In the case of cold plasma, there are two types of O-modes: superluminal, with $\omega>k_\parallel c$, and subluminal, with $\omega<k_\parallel c$. In fact, the usual MHD Alfvén wave is the low-frequency limit of the subluminal O-mode. 

We assume that the distribution function can be approximately considered one-dimensional due to strong synchrotron losses. Therefore, $f_s(\mathbf{p})=(1/p_\perp)f_{\parallel,s}(\mathcal{E})\delta(p_\perp)$, where $\mathcal{E}=|p_z|c$ is the longitudinal energy, and
\begin{equation}
  \int f_s(\mathbf{p})\text{d}\mathbf{p}=\frac{2}{c}\int\limits f_{\parallel,s} (\mathcal{E})\text{d}\mathcal{E}=n_0.
\end{equation}
In this case, the dispersion equation for the O-mode in the hot strongly magnetized plasma (e.g. \citealt{Arons1986}, \citealt{Gedalin1998}) is
\begin{equation}
    k_\perp^2+\left(k_\parallel^2-\frac{\omega^2}{c^2}\right)\left(1+\frac{4\pi\text{i}}{\omega}\sigma(\omega,k_\parallel)\right)=0,
\end{equation}
where
\begin{equation}
\sigma(\omega,k_\parallel)=\text{i}e^2\sum_s\int\frac{(\partial f_{\parallel,s}/\partial p_z)\text{d}p_z}{\omega-k_\parallel v_z+\text{i}0}=\frac{\text{i}e^2\omega}{m}\sum_s\int\frac{f_{\parallel,s}\text{d}p_z}{\Gamma^3(\omega-k_\parallel v_z)^2}.
\end{equation}
Let us assume that Landau damping is weak. This limit is important for our analysis, since we are interested in the small correction to the MHD dispersion relation for Alfvén waves. Therefore, one can write (see also \citealt{Vega2024})
\begin{equation}\label{con}
    \Big|1-\frac{\omega}{k_\parallel c}\Big|\ll 
    \begin{cases}
    1/(2\Gamma_\text{max}^2), \,\,\, \text{if} \,\, \alpha<2, \\
    1/(2\Gamma_\text{min}^2), \,\,\, \text{if} \,\, \alpha\geq 2.
    \end{cases}
\end{equation}
It leads to the following expression for the conductivity
\begin{equation}
    \sigma(\omega,k_\parallel)\approx \frac{\text{i}e^2\omega}{m k_\parallel^2c^2}\sum_s\int \frac{f_{\parallel,s}\text{d}p_z}{\Gamma^3(1-v_z/c)^2}=\frac{8\text{i}n_0e^2\omega}{m k_\parallel^2 c^2}\langle \Gamma\rangle.
\end{equation}
The solution of the dispersion equation is
\begin{equation}\label{d}
    \omega^2\approx k_\parallel^2c^2\left(1-\frac{k_\perp^2 d^2}{4\langle\Gamma\rangle^2}\right),
\end{equation}
where $d^2=c^2\langle\Gamma\rangle/(\omega^2_{pe})$, $\omega_{pe}=\sqrt{8\pi n_0 e^2/m}$ and the dispersion relation is written in such a form that it is more convenient to check the condition~(\ref{con}). If $k_\perp^2d^2\ll 1$, the inequality~(\ref{con}) is fulfilled and the MHD approximation for Alfvén waves can be used. Therefore, the characteristic scale at which the dissipative interval begins and Landau damping becomes important is $d$. However, unlike in the isotropic case, only the longitudinal energy needs to be averaged.

\bibliography{References}{}
\bibliographystyle{aasjournal}



\end{document}